\documentclass[a4paper,conference,onecolumn]{IEEEtran}

\usepackage[utf8]{inputenc}
\usepackage[T1]{fontenc}
\usepackage{url}
\usepackage{ifthen}
\usepackage{amsfonts}
\usepackage{cite}
\usepackage{tikz}
\usepackage{cancel}
\usepackage{xcolor}
\usepackage{float}
\usepackage{bm}
\usepackage{hyperref}
\hypersetup{pdfborder = 0 0 0}

\usepackage{amsthm}

\newtheorem{theorem}{Theorem}

\newtheorem{proposition}{Proposition}

\theoremstyle{definition}
\newtheorem{definition}{Definition}
\usepackage{amssymb}

\usetikzlibrary{patterns}
\usetikzlibrary{shapes,arrows}
\newcommand{\Exp}[2][]{\ensuremath{\mathbb{E}_{#1}\left[#2 \right]}} 

\newcommand{\eqann}[2][=]{\overset{\mathclap{(\text{#2})}}{#1}} 
\newcommand{\eqannref}[1]{$(\text{#1})$}
\newcommand{\bv}[1]{\mathbf{#1}} 
\newcommand{\rv}[1]{\mathsf{#1}} 
\newcommand{\Prob}[1]{\ensuremath{\mathbb{P} \left\{#1 \right\}}} 

\newcounter{tempEquationCounter} 
\newcounter{thisEquationNumber}

\usepackage[cmex10]{mathtools} 
\renewcommand{\Pr}{\mathbb{P}}

\begin{document}
\title{Broadcasting Information subject to State Masking }



 \author{%
   \IEEEauthorblockN{Michael Dikshtein \href{https://orcid.org/0000-0002-0498-273X}{\includegraphics[width=0.32cm]{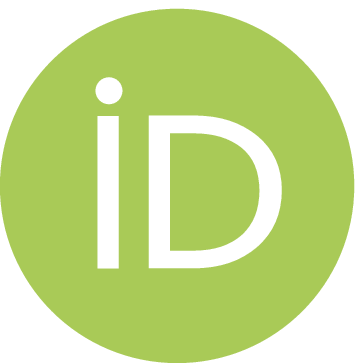}},
   Shlomo Shamai (Shitz)}
   \IEEEauthorblockA{%
              Department of EE, 
              Technion, 
              Haifa 32000, Israel,
              \{michaeldic@campus,sshlomo@ee\}.technion.ac.il}
 }
%

\maketitle

\begin{abstract}
	We study the problem of coding over a general discrete memoryless broadcast channel controlled by random parameters. The parameters are available at the transmitter in a non-causal manner and are subject to a state masking constraint on the receivers. We derive inner and outer bounds on the achievable region and show that for the special case of Gaussian broadcast channel with private messages, these bounds are tight.
\end{abstract}


\section{Introduction}

\begin{figure}
	\centering
	\tikzstyle{block} = [draw, rectangle, minimum width = 0.5cm, minimum height = 0.5cm]
	\tikzstyle{sum} = [draw,circle]
	\tikzstyle{input} = [coordinate]
	\tikzstyle{dummy} = [coordinate]
	\tikzstyle{output} = [coordinate]
	\tikzstyle{amp} = [draw,shape border rotate = 180,regular polygon,regular polygon sides=3]
	
	\begin{tikzpicture}
	
	\node [input]	(in)	at 	(0,0) 		{};
	\node [block]	(enc) 	at 	(1,0)		{\tiny{Enc}};
	\node [block]	(state) at 	(2.5,1.5)		{\tiny{$ P_{\rv{S}} $}};
	\node [draw, rectangle, minimum width = 1cm, minimum height = 1.5cm]	(ch)	at 	(2.5,0)		{\tiny{$P_{\rv{Y}_1 \rv{Y}_2|\rv{X} \rv{S}}$}};
	\node [block]	(dec1)	at	(4,0.5)		{\tiny{Dec 1}};
	\node [block]	(dec2)	at	(4,-0.5)		{\tiny{Dec 2}};
	\node [output]	(out1)	at	(5,0.5)		{};
	\node [output]	(out2)	at	(5,-0.5)		{};
	
	\draw [->] (in) node[left] {\tiny{$\rv{M}_0, \rv{M}_1, \rv{M}_2$}} --  (enc);
	\draw [->] (state) -- (state-|enc) -- node [right] {\tiny{$ \rv{S}^n $}} (enc) ;
	\draw [->] (state) -- node [right] {\tiny{$ \rv{S}^n $}} (ch);
	\draw [->] (enc) -- node[above] {\tiny{$\rv{X}^n$}} (ch);
	\draw [->] (ch.40) --  node[above] {\tiny{$\rv{Y}_1^n$}} (dec1)  ;
	\draw [->] (ch.320) --  node[above] {\tiny{$\rv{Y}_2^n$}} (dec2)  ;
	\draw [->] (dec1) --  (out1);
	\draw [->] (dec2) --  (out2);
	\node at (5.5,0.7) {\tiny{$\hat{\rv{M}}_0,\hat{\rv{M}}_1$}};
	\node at (6,0.3) {\tiny{$E_{1}\geq \frac{1}{n} I(\rv{S}^n;\rv{Y}_1^n)$}};
	\node at (5.5,-0.3) {\tiny{$\hat{\rv{M}}_0,\hat{\rv{M}}_2$}};
	\node at (6,-0.7) {\tiny{$E_{2}\geq \frac{1}{n} I(\rv{S}^n;\rv{Y}_2^n)$}};
	\end{tikzpicture}
	\caption{System model for general BC subject to state masking constraints.}
	\label{figure:system_model}
\end{figure}
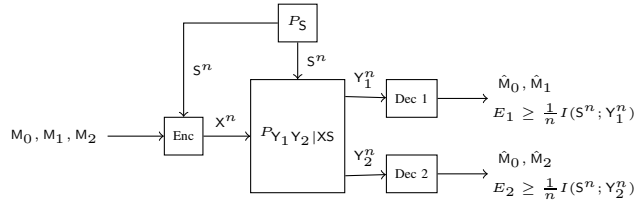

We consider a discrete memoryless broadcast channel (DMBC) with random parameters and channel side information (CSI) known in a noncausal manner to the transmitter subject to a state masking criterion at the receivers, depicted in Figure \ref{figure:system_model}.

The single-letter expression for the capacity of the point to point discrete memoryless channel (DMC) with noncausal CSI at the encoder (the G-P channel) was derived in the seminal work of Gel'fand and Pinsker \cite{gelf80}. One of the most interesting special cases of the G-P channel is the Gaussian additive noise and interference setting in which the additive interference plays the role of the state sequence, which is known non-causally to the transmitter. Costa showed in \cite{Costa83} that the capacity of this channel is equal to the capacity of the same channel without additive interference. The capacity achieving scheme of \cite{Costa83} (which is that of \cite{gelf80} applied to the Gaussian case) is termed ``writing on dirty paper" (WDP). Cohen and Lapidoth \cite{cohen2002gwdp} showed that any interference sequence can be totally removed when the channel noise is ergodic and Gaussian.

The DMBC was first introduced by Cover \cite{cover1972}. The capacity region of the DMBC is still an open problem. The largest known inner bound on the capacity region of the DMBC with private messages was derived by Marton \cite{marton1979}. Liang \cite{liang2005} derived an inner bound on the capacity region of the DMBC with an additional common message. The best outer bound for DMBC with a common message is due to Nair and El Gamal \cite{nair2006}. There are however some special cases where the capacity region is fully characterized. For example the capacity region of the degraded DMBC was established by Gallager \cite{gallager1974}. The capacity region of the Gaussian BC was derived by Bergmans \cite{bergmans1974}. An interesting result is the capacity region of the Gaussian MIMO BC which was established by Weingarten et al. \cite{weingarten2006}. The authors introduced a new notion of \textit{an enhanced channel} and used it jointly with the Entropy Power Inequality (EPI) to show their result. The capacity achieving scheme relies on the dirty paper coding technique.  Liu and Viswanath \cite{liu2007} developed \textit{an extremal inequality} proof technique and showed that it can be used to establish a converse result in various Gaussian MIMO multiterminal networks, including the Gaussian MIMO BC with private messages. Recently, Geng and Nair \cite{geng2014} developed a different technique to characterize the capacity region of Gaussian MIMO BC with common and private messages.

Degraded DMBC with causal and noncausal side information was introduced by Steinberg \cite{steinberg2005}. Inner and outer bounds were derived on the capacity region. For the special case in which the nondegraded user is informed about the channel parameters, it was shown that the bounds are tight, thus deriving the capacity region for that case. The general DMBC with noncausal CSI at the encoder was studied by Steinberg and Shamai \cite{steinbergshamai2005}. An inner bound was derived and it was shown to be tight for the Gaussian BC with independent additive interference at both channels. Outer bounds for DMBC with CSI at the encoder were derived in \cite{khosravi2011}.

The problem of state-masking and information rate trade-off was introduced in \cite{merhav2007}. In that work, the state sequence was treated as an undesired information that leaks to the receiver and is known to the transmitter. The measure of ability of the receiver to learn about the state from the received sequence was defined as the normalized block-wise mutual information between the state sequence $ \rv{S}^n $ and the received sequence $ \rv{Y}^n $, that is, $ I(\rv{S}^n;\rv{Y}^n)/n$.

The concept of state amplification is a dual problem to state masking. Kim et al. \cite{kim2008} considered the problem of transmitting data at rate $ R $ over a DMC with random parameters and CSI at the encoder and simultaneously conveying the information about the channel state itself to the receiver.  They defined the channel state uncertainty reduction rate to be $ \Delta \triangleq \frac{1}{n} (H(S^n)-\log |L_n|) $, where $ |L_n| $ is the receiver list size in list decoding of the state, and found the $ (R,\Delta) $ achievable region. 

Courtade \cite{courtade2012} considered a joint scenario, with two-encoder source coding setting where one source is to be amplified, while the other source is to be masked. Koyluoglu et al. \cite{koyluoglu2016} considered a state-dependent BC with state sequence known in noncausal manner to Alice (the transmitter) and its goal is to effectively convey the state to Bob (receiver 1) while "masking" it from Eve (receiver 2). Liu and Chen \cite{liu2009} considered the problem of message transmission and state estimation over Gaussian BC, where both received signals interfered by same additive Gaussian state.
Grover and Sahai \cite{grover2010} related the problem of state masking to Witsenhausen's Counter-example \cite{witsenhausen1968counterexample}. Tutuncuoglu et al. \cite{tutuncuoglu2014} studied the problem of state amplification and state masking in an energy harvesting binary channel. They considered a situation where the binary encoder is connected to a battery source $ \rv{B}_i $ which tries to harvest energy $ \rv{E}_i $ at every time slot $ i $ and were interested in how much the decoder, that has no knowledge of the battery state $ \rv{B}_i $ nor the energy process $ \rv{E}_i $, can learn about the energy arrival process $ \rv{E}^n $. A privacy-constrained information extraction problem was recently considered by Asoodeh et al. \cite{asoodeh2016}. In their setting, they divided the information to be conveyed into private information and public information and also used the mutual information measures to determine the trade-off between public information transmission and private data leakage. A good tutorial on channel coding in the presence of CSI that also covers the state masking setting can be found in \cite{keshet2008channel}.

In this work, we extend the state masking scenario to a broadcast channel corrupted by state which is known non-causally to the encoder. In our setting the encoder wishes to reliably transmit common and private information over state-dependent channel to two receivers, while simultaneously minimizing the amount of information each receiver can learn about the state sequence $ s^n $. We develop inner and outer bounds and show that they are tight for a special case of state-dependent Gaussian BC with private messages.

\section{Notations and Problem Formulation}

Throughout the paper, random variables are denoted using a sans-serif font, e.g., $ \rv{X} $, their realizations are denoted by the respective lower case letters, e.g., $ x $, and their alphabets are denoted by the respective calligraphic letter, e.g., $ \mathcal{X} $. Let $ \mathcal{X}^n $ stand for the set of all $ n $-tuples of elements from $ \mathcal{X} $. An element from $ \mathcal{X}^n $ is denoted by $ x^n = (x_1,x_2,\dots , x_n) $ and substrings by $ x_{i}^j = (x_i,x_{i+1},\dots ,x_j) $. The cardinality of a finite set, say $ \mathcal{X}$, is denoted by $ |\mathcal{X}|$. The probability distribution function of $ \rv{X} $, the joint distribution function of $ \rv{X} $ and $ \rv{Y} $, and the conditional distribution of $ \rv{X} $ given $ \rv{Y} $ are denoted by $ P_\rv{X} $, $ P_{\rv{X},\rv{Y}} $ and $ P_{\rv{X}|\rv{Y}} $ respectively. The expectation of $ \rv{X} $ is denoted by $ \Exp{\rv{X}} $. The probability of an event $ \mathcal{E} $ is denoted as $ \Prob{\mathcal{E}} $. The set of jointly $ \epsilon $-typical $ n $-tuples $ (x^n,y^n) $ is defined as $ \mathcal{T}_\epsilon^{(n)} (P_{\rv{X}\rv{Y}}) $ \cite{gamal2011network}.

A set of consecutive integers starting at $ 1 $ and ending in $ 2^{nR} $ are denoted as $ \mathcal{I}_{R} \triangleq \{1,2,\dots, 2^{nR} \} $.

Let $ \mathcal{X},\mathcal{S},\mathcal{Y}_1,\mathcal{Y}_2 $ be finite sets, and let $ P_S $ be a probability mass function (pmf) on $ \mathcal{S} $. We consider a 2-receiver discrete memoryless broadcast channel with random parameters $ (\mathcal{S},P_\rv{S},\mathcal{X}, P_{\rv{Y}_1,\rv{Y}_2|\rv{X},\rv{S}}, \mathcal{Y}_1 \times \mathcal{Y}_2) $ that consists of an input alphabet $ \mathcal{X} $, a state alphabet $ \mathcal{S} $ and two output alphabets $ \mathcal{Y}_1$ and $ \mathcal{Y}_2 $  and a probability transition function $ P_{\rv{Y}_1,\rv{Y}_2|\rv{X},\rv{S}} $, where the states $ \rv{S}_i $, $ i=1,2,\dots $, are random taking values in $ \mathcal{S} $ and drawn from a discrete memoryless source (DMS) $ P_{\rv{S}^n} (s^n) = \prod_{i=1}^{n} P_\rv{S}(s_i)$.
The channel is assumed to be memoryless and without feedback. Thus, probabilities on $ n $-tuples are given by:
\begin{equation*}
	P_{\rv{Y}_1^n\rv{Y}_2^n|\rv{X}^n\rv{S}^n} (y_1^n,y_2^n|x^n,s^n) = \prod_{i=1}^{n} P_{\rv{Y}_1\rv{Y}_2|\rv{X}\rv{S}} (y_{1i},y_{2i}|x_i,s_i)	.
\end{equation*}
The channel input signal is subject to an average input cost constraint $ \frac{1}{n} \sum_{i=1}^{n} \phi(\rv{X_i}) \leq \Gamma $, where $ \phi: \mathcal{X} \rightarrow \mathbb{R}^+ $ is the input cost function and $ \Gamma>0 $ is a given constant.

A $(2^{nR_0},2^{nR_1},2^{nR_2},n) $ code for the broadcast channel with state sequence $ \rv{S}^n $ known non-causally at the encoder consists of
\begin{itemize}
	\item Three message sets $ \mathcal{I}_{R_0} $, $ \mathcal{I}_{R_1} $ and $ \mathcal{I}_{R_2} $.
	\item An encoder that assigns a codeword $ x^n(m_0,m_1,m_2,s^n) $ to each message-state quadruple $ (m_0,m_1,m_2,s^n) \in \mathcal{I}_{R_0} \times \mathcal{I}_{R_1}\times \mathcal{I}_{R_2} \times \mathcal{S}^n $.
	\item Two decoders, where decoder 1 assigns an estimate  $ \hat{m}_{01} \in \mathcal{I}_{R_0} $ and $ \hat{m}_1 \in \mathcal{I}_{R_1} $ to each received sequence $ y_1^n $, and decoder 2 assigns an estimate $ \hat{m}_{02} \in \mathcal{I}_{R_0} $ and $ \hat{m_2} \in \mathcal{I}_{R_2} $ to each received sequence $ y_2^n $.
\end{itemize}
Let $(\hat{\rv{M}}_{01},\hat{\rv{M}}_1) $ and $(\hat{\rv{M}}_{02},\hat{\rv{M}}_2)$ denote the outputs of decoder $1$ and decoder $2$, respectively.  We assume that the message triple $ (\rv{M}_0, \rv{M}_1, \rv{M}_2) $ is uniformly distributed over $ \mathcal{I}_{R_0} \times \mathcal{I}_{R_1} \times \mathcal{I}_{R_2} $. The average probability of error is defined as
\begin{equation}
	P_e^{(n)} =	\Prob{\bigcup_{k=1}^2 \{(\hat{\rv{M}}_{0k},\hat{\rv{M}}_{k}) \neq (\rv{M}_0 ,\rv{M}_k) \} }.
\end{equation}
The average probability of error at each receiver is defined as 
\begin{equation}
	P_{e,k}^{(n)} = \Prob{(\hat{\rv{M}}_{0k} , \hat{\rv{M}}_k) \neq (\rv{M}_0,\rv{M}_k)}, \quad k=1,2 .
\end{equation}
Obviously the average probability $ P_e^{(n)} $ tends to zero as $ n \rightarrow \infty $ , iff both $ P_{e,1}^{(n)} $ and $ P_{e,2}^{(n)} $ tend to zero as $ n \rightarrow \infty $.

We are interested in the interplay between reliable coding at rate triples $ (R_0,R_1,R_2) $ which we would like to keep as high as possible and the (normalized) mutual informations $ I(\rv{S}^n;\rv{Y}_1^n)/n $ and $ I(\rv{S}^n;\rv{Y}_2^n)/n $, which we would like to make as small as possible. 
\begin{definition}
	
For a given $ \Gamma >0 $, a quintuple $ (R_0,R_1,R_2,E_{1},E_{2}) $ is said to be achievable if for every $ \epsilon>0 $ and sufficiently large $ n $, there exists a sequence of $ (2^{nR_0},2^{nR_1},2^{nR_2},n) $ codes such that the following conditions are simultaneously satisfied:
\begin{subequations}
	\begin{align}
			\frac{1}{n} \sum_{i=1}^{n} \phi(X_i) &\leq \Gamma,\\
		P_e^{(n)} &\leq \epsilon,\\
	 	\frac{1}{n} I(\rv{S}^n;\rv{Y}_k^n) & \leq E_{k}+\epsilon, \quad  k=1,2.
 	\end{align}
\end{subequations}
\end{definition}
\begin{definition}
	The achievable region $ \mathcal{R} $ is the closure of the set of all achievable quintuples $ \{(R_0,R_1,R_2,E_{1},E_{2})\} $ .

\end{definition}

\begin{definition}
	The achievable region $ \mathcal{R}_0 $ is the set of all zero-rates achievable pairs $ \{(R_0,R_1,R_2,E_{1},E_{2})\}=\{(0,0,0,E_{1},E_{2})\} $ .
	
\end{definition}

\section{Main Results}
As mentioned before, the capacity region of the general DMBC is unknown even for channels without state. In this section we present inner and outer bounds on the achievable region. We begin with the inner bound. The inner bound on the rate-triple $ (R_0,R_1,R_2) $ is essentially the same as was given in \cite{steinbergshamai2005}, but our proof is simpler, and we also contribute a bound on the equivocation rate-pair $ (E_1,E_2) $. The main idea behind the proof is integration of Marton and GP coding, where for each message, a subcodebook is generated, whose size is large enough such that for every state sequence $ s^n $ a jointly typical auxiliary codeword can be found in the subcodebook.
\begin{proposition} \label{proposition:inner_bound_positive_rates}
	An achievable region $ \mathcal{R} $ consists of a quintuple $ (R_0,R_1,R_2,E_1,E_2) $ that satisfies the following conditions
	\begin{subequations} \label{eq:SDDMBC_inner_bound}
		\begin{align}
		R_0 \leq \min& \{ I(\rv{W};\rv{Y}_1), I(\rv{W};\rv{Y}_2)\}-I(\rv{W};\rv{S}),\\
		R_0+R_1 &\leq I(\rv{W},\rv{U};\rv{Y}_1)-I(\rv{W},\rv{U};\rv{S}),\\
		R_0+R_2 &\leq I(\rv{W},\rv{V};\rv{Y}_2)-I(\rv{W},\rv{V};\rv{S}),\\
		R_0+R_1+R_2 &\leq I(\rv{W},\rv{U};\rv{Y}_1)-I(\rv{W},\rv{U};\rv{S}) + I(\rv{W},\rv{V};\rv{Y}_2)-I(\rv{W},\rv{V};\rv{S}) \\
		&\quad-\min \{ I(\rv{W};\rv{Y}_1), I(\rv{W};\rv{Y}_2)\} -I(\rv{W};\rv{S}) - I(\rv{U};\rv{V}|\rv{W},\rv{S}) ,\\
		E_1 &\leq I(\rv{S};\rv{W},\rv{U},\rv{Y}_1), \\
		E_2 &\leq  I(\rv{S};\rv{W},\rv{V},\rv{Y}_2),
		\end{align}
	\end{subequations}
	for some pmf $ P_{\rv{S}\rv{W}\rv{U}\rv{X}\rv{Y}_1\rv{Y}_2}=P_{\rv{S}}P_{\rv{W}\rv{U}\rv{V}\rv{X}|\rv{S}} P_{\rv{Y}_!\rv{Y}_2|\rv{X}\rv{S}} $.
\end{proposition}
We give an outline of the proof in Section \ref{proof_outline:inner_bound} while the full proof is relegated to Appendix \ref{section:inner_bound_proof}.

Next, we provide the outer bound on $ \mathcal{R} $.
\begin{proposition}	\label{proposition:outer_bound_positive_rates}
	If a rate quintuple $ (R_0,R_1,R_2,E_1,E_2) $ is achievable for the DM-BC with random parameters and CSI known non-causally at the transmitter, then there exists a distribution $P_{\rv{W}\rv{U}\rv{V}\rv{X}|\rv{S}}$ such that the following inequalities are satisfied:
	\begin{subequations} \label{eq:SDDMBC_outer_bound}
		\begin{align}
		R_0 &\leq \min \{ I(\rv{W};\rv{Y}_1|\rv{S}), I(\rv{W};\rv{Y}_2|\rv{S})\} \\
		R_0+R_1 &\leq \min \{ I(\rv{W};\rv{Y}_1|\rv{S}), I(\rv{W};\rv{Y}_2|\rv{S})\}
		+I(\rv{U};\rv{Y}_1|\rv{W},\rv{S}) \\
		R_0+R_2 &\leq \min \{ I(\rv{W};\rv{Y}_1|\rv{S}), I(\rv{W};\rv{Y}_2|\rv{S})\} + I(\rv{V};\rv{Y}_2|\rv{W},\rv{S}) \\
		R_0+R_1+R_2 &\leq \min \{ I(\rv{W};\rv{Y}_1|\rv{S}), I(\rv{W};\rv{Y}_2|\rv{S})\} 
		+ I(\rv{U};\rv{Y}_1|\rv{W},\rv{S})+I(\rv{X};\rv{Y}_2|\rv{W},\rv{U},\rv{S}) \\
		R_0+R_1+R_2 &\leq \min \{ I(\rv{W};\rv{Y}_1|\rv{S}), I(\rv{W};\rv{Y}_2|\rv{S})\} 
		 + I(\rv{X};\rv{Y}_1|\rv{W},\rv{V},\rv{S})+I(\rv{V};\rv{Y}_2|\rv{W},\rv{S}) \\
		E_k &\geq I(\rv{S};\rv{Y}_k) \quad k=1,2,
		\end{align}
	\end{subequations}
where $P_{\rv{S}\rv{W}\rv{U}\rv{V}\rv{X}\rv{Y}_1\rv{Y}_2}=P_{\rv{S}} P_{\rv{W}\rv{U}\rv{V}\rv{X}|\rv{S}}P_{\rv{Y}_1\rv{Y}_2|\rv{X}\rv{S}}$.
\end{proposition}
We give an outline of the proof in Section \ref{proof_outline:outer_bound} while the full proof can be found in Appendix \ref{appendix:outer_bound_common}.

In some practical applications the transmitter is solely intended to minimize the leakage without transmitting any information. This scenario is also  simpler to analyze in the zero rate triple case $ (R_0,R_1,R_2)=(0,0,0) $. In such a scenario our goal is to minimize $ I(S^n;Y_1^n)/n $ and $ I(S^n;Y_2^n)/n $.

Let $ \mathcal{E}_0 (\rv{X}) $ be the set of equivocation rate pairs $ (E_1,E_2) $ such that $E_k \geq I(\rv{S};\rv{Y}_k) $, $ k=1,2$.
Following is a characterization of the achievability region for $(R_0,R_1,R_2)=(0,0,0)$. 
\begin{theorem} \label{thrm:dmc_bc_zero_rate}
	The achievable zero-rates region $ \mathcal{R}_0 $ of the DM-BC with random parameters $ p(y_1,y_2|x,s) $ is the convex hull of the union of the regions $ \mathcal{E}_0 (\rv{X}) $ over all $ p(x|s) $.
\end{theorem}
\begin{IEEEproof}
	The theorem follows from the inner bound in Proposition \ref{proposition:inner_bound_positive_rates} and the outer bound in  Proposition \ref{proposition:outer_bound_positive_rates}, and respectively \eqref{eq:SDDMBC_inner_bound} and \eqref{eq:SDDMBC_outer_bound} by the following choice of auxiliary random variables: $ \rv{W}= \emptyset $, $ \rv{U}=\emptyset $ and $ \rv{V} = \emptyset$. In this case, the encoder simply generates $\rv{X}^n$ given $\rv{S}^n$ according to $\prod_{i=1}^n P_{\rv{X}|\rv{S}}(x_i|s_i)$. Since this creates a memoryless "channel" from $\rv{S}$ to $(\rv{Y}_1,\rv{Y}_2)$ we get that $I(\rv{S}^n;\rv{Y}_k^n)/n=I(\rv{S};\rv{Y})$. 
\end{IEEEproof}
\section{Proofs Outline}
In this section we provide an outline to the proofs of Proposition \ref{proposition:inner_bound_positive_rates} and Proposition \ref{proposition:outer_bound_positive_rates}.
\subsection{Inner Bound} \label{proof_outline:inner_bound}
Fix the conditional pmf $ P_{\rv{W}\rv{U}\rv{V} \rv{X}|\rv{S}} $ and let $ n\rightarrow \infty $. Randomly and independently generate $ 2^{n(R_0+\tilde{R}_0)} $ sequences $ w^n(m_0,l_0) $, $ m_0 \in \mathcal{I}_{R_0} $, $ l_0 \in \mathcal{I}_{\tilde{R}_0} $, according to $\prod_{i=1}^n P_{\rv{W}}(w_i)$. For each $ (m_0,l_0) $, generate $ 2^{n(R_2+\tilde{R}_2)} $ independent sequences $ v^n(m_0,l_0,m_2,l_2) $, $ m_2 \in \mathcal{I}_{R_2} $, $ l_2\in \mathcal{I}_{\tilde{R}_2} $, according to $ \prod_{i=1}^n P_{\rv{V}|\rv{W}} (v_i|w_i(m_0,l_0)) $. Similarly, for each $ (m_0,l_0) $, generate $ 2^{n(R_1+\tilde{R}_{1s}+\tilde{R}_{12})} $ independent sequences $u^n(m_0,l_0,m_1,l_{1s},l_{12}) $, $ m_1 \in \mathcal{I}_{R_1} $, $ l_{1s} \in \mathcal{I}_{\tilde{R}_{1s}} $, $ l_{12} \in \mathcal{I}_{\tilde{R}_{12}} $, according to $ \prod_{i=1}^n P_{\rv{U}|\rv{W}} (u_i|w_i(m_0,l_0))$. Let $ (m'_0,m'_1,m'_2) $ be the message triple to be sent with the state sequence $ s^n $ observed. First the encoder finds $ \tilde{l}_0 $, such that $ (s^n,w^n(m'_0,\tilde{l}_0)) \in \mathcal{T}^{(n)}_{\epsilon'} $. It can be shown that at least one such $ \tilde{l}_0 $ exists if $ \tilde{R}_0>I(\rv{W};\rv{S}) $. Then, given $ w^n(m'_0,\tilde{l}_0) $, the encoder finds $ \tilde{l}_2 $, such that $ (s^n,w^n(m'_0,\tilde{l}_0),v^n(m'_0,\tilde{l}_0,m'_2,\tilde{l}_2)) \in \mathcal{T}_{\epsilon''}^{(n)} $.  It can be shown that at least one such $ \tilde{l}_2 $ exists if $ \tilde{R}_2>I(\rv{V};\rv{S}|\rv{W}) $. Similarly, given $ w^n(m'_0,\tilde{l}_0) $, the encoder finds $ \tilde{l}_{1s} $, such that $ (s^n,w^n(m'_0,\tilde{l}_0),u^n(m'_0,\tilde{l}_0,m'_1,\tilde{l}_{1s},l_{12})) \in \mathcal{T}_{\epsilon''}^{(n)} $ for every $ l_{12} $. It can be shown that at least one such $ \tilde{l}_{1s} $ exists if $ \tilde{R}_{1s}>I(\rv{U};\rv{S}|\rv{W}) $. Then, given $ w^n(m'_0,\tilde{l}_0) $, $ v^n(m'_0,\tilde{l}_0,m'_2,\tilde{l}_2) $ and $ \tilde{l}_{1s} $, the encoder finds $ \tilde{l}_{12} $, such that $ (s^n,w^n(m'_0,\tilde{l}_0),v^n(m'_0,\tilde{l}_0,m'_2,\tilde{l}_2), u^n(m'_0,\tilde{l}_0,m'_1,\tilde{l}_{1s},\tilde{l}_{12})) \in \mathcal{T}_{\epsilon'''}^{(n)} $. It can be shown that at least one such $ \tilde{l}_{12} $ exists if $ \tilde{R}_{1s}>I(\rv{U};\rv{V}|\rv{W},\rv{S}) $. Finally, for each quadruple $ (m_0,m_1,m_2,s^n) $ generate a sequence $ x^n(m_0,m_1,m_2,s^n) $ according to $ \prod_{i=1}^{n} P_{\rv{X}|\rv{W}\rv{U}\rv{V}\rv{S}} (x_i|w_i,u_i,v_i) $. In order to transmit $ (m_0,m_1,m_2) $ given $ s^n $ send $ x^n(m_0,m_1,m_2,s^n) $.

Decoders $ 1 $ and $ 2 $ use joint typicality decoding of $ (w^n,u^n,y_1^n) $ and $ (w^n,v^n,y_2^n) $ respectively. It can be shown with probability approaching 1 as $ n \rightarrow \infty  $ the following rates are achievable
	\begin{equation}
	\begin{split}
	R_0 &\leq \min \{I(\rv{W};\rv{Y}_1),I(\rv{W};\rv{Y}_2)\}-I(\rv{W};\rv{S}) ,\\
	R_0+R_2 &\leq I(\rv{W},\rv{V};\rv{Y}_2)-I(\rv{W},\rv{V};\rv{S}), \\
	R_0+R_1 &\leq I(\rv{W},\rv{U};\rv{Y}_1)-I(\rv{W};\rv{S})-I(\rv{U};\rv{V},\rv{S}|\rv{W}) \\
	&= I(\rv{W},\rv{U};\rv{Y}_1)-I(\rv{W},\rv{U};\rv{S})-I(\rv{U};\rv{V}|\rv{W},\rv{S}).
	\end{split}
	\end{equation}
	
	As for the upper bound on the mutual information between $ \rv{S}^n $ and $ \rv{Y}_1^n $,
	\begin{equation} \label{eq:E1_direct_1}
	\begin{split}
	I(\rv{S}^n;\rv{Y}_1^n) &\leq I(\rv{S}^n;\rv{W}^n,\rv{U}^n,\rv{Y}_1^n) \\
	&\leq I(\rv{S}^n;\rv{W}^n,\rv{U}^n|\rv{M}_0,\rv{M}_1)+I(\rv{S}^n;\rv{Y}_1^n|\rv{W}^n,\rv{U}^n ) \\
	&\eqann[\leq]{a} H(\rv{W}^n|\rv{M}_0)+H(\rv{U}^n|\rv{W}^n,\rv{M}_0,\rv{M}_1)
	-H(\rv{U}^n|\rv{W}^n,\rv{M}_0,\rv{M}_1,\rv{S}^n)+nI(\rv{S};\rv{Y}_1|\rv{W},\rv{U} ) \\
	&\eqann[\leq]{b} n( \tilde{R}_0+\tilde{R}_{1s}+\tilde{R}_{12}-\tilde{R}_{12}+I(\rv{S};\rv{Y}_1|\rv{W},\rv{U} )) \\
	&= n I(\rv{S};\rv{Y}_1,\rv{W},\rv{U})
	\end{split}
	\end{equation}
	where \eqannref{a} follows from  the memorylessness of the channel $ P_{\rv{Y}_2|\rv{W},\rv{U},\rv{S}} $. In \eqannref{b} we used the fact that the sizes of each bin of $ \mathcal{C}_0 $ and $ \mathcal{C}_1 $  are $ 2^{n\tilde{R}_0} $ and $ 2^{n(\tilde{R}_{1s}+\tilde{R}_{12})} $, respectively. Furthermore, given $ (w^n(m_0,l_0),m_0,m_1,s^n) $, $ u^n $ is uniform over $ \mathcal{I}_{\tilde{R}_{12}}$.
	The upper bound for $ I(\rv{S}^n;\rv{Y}_2^n) $ follows from similar considerations.
\subsection{Outer bound} \label{proof_outline:outer_bound}
The outer bound on the achievable rates region can be shown by providing the state sequence $ s^n $ as side information to the receivers, defining the following auxiliary random variables for each $ i\in [1:n] $
\begin{equation}
\rv{W}_{i} \triangleq (\rv{M}_0,\rv{Y}_{1}^{i-1}, \rv{S}^{i-1},\rv{Y}_{2,i+1}^n,\rv{S}_{i+1}^n), \quad \rv{U}_i=\rv{M}_1,\quad \rv{V}_i \triangleq \rv{M}_2,
\end{equation}
and a proper use of Csisz\'ar and K\"{o}rner sum identity \cite{csiszar1978broadcast}.

As for the lower bound on the equivocation rates $ E_k $, $ k=1,2 $, we use the memorylessness property of the source $ P_{\rv{S}} $ to show
\begin{equation}
I(\rv{S}^n;\rv{Y}_k^n) \geq \sum_{i=1}^n I(\rv{S}_i;\rv{Y}_{k,i}) \geq n I(\rv{S};\rv{Y}_{k}).
\end{equation}
\section{State-Dependent Gaussian BC}
In this section we consider a scalar additive white Gaussian noise BC with additive state. The channel outputs corresponding to the inputs $ (\rv{X},\rv{S}_1,\rv{S}_2) $ are:
\begin{equation}
	\rv{Y}_k = \rv{X}+\rv{S}_k+\rv{Z}_k, \quad k=1,2
\end{equation}
where $ \rv{Z}_k \sim \mathcal{N} (0,N_k) $ , $ k\in \{1,2\}$ are additive Gaussian noises, $ \rv{S}_k \sim \mathcal{N} (0,Q_k), k\in \{1,2\} $ are additive Gaussian random variables, both known noncausally at the transmitter. The Gaussian random variables $ \rv{Z}_1,\rv{Z}_2,\rv{S}_1 $ and $ \rv{S}_2 $ are mutually independent and the equivocation rates are measured between the outputs and the state $\bv{S}=(\rv{S}_1,\rv{S}_2)$. The input $ \rv{X} $ is power constrained to $ P $, such that, $ \frac{1}{n} \sum_{i=1}^{n}\rv{X}_i^2 \leq P $ . We further assume that $ N_2>N_1 $ without loss of generality.
Denote $ P' \triangleq (1-\rho_1^2-\rho_2^2) P $.
\begin{theorem}
	The rate-leakage region of the Gaussian State-Dependent Broadcast Channel with private messages is the quadruple $ (R_1,R_2,E_{1},E_{2} ) $ such that
	\begin{align}
	R_1 &\leq \frac{1}{2} \log \left(1+ \frac{\gamma  P'}{N_1} \right), \label{eq:R1_Scalar_Gaussian_Private_Capacity}\\
	R_2 &\leq \frac{1}{2} \log \left(1+\frac{ \overline{\gamma}  P'}{\gamma  P'+N_2} \right),\label{eq:R2_Scalar_Gaussian_Private_Capacity}\\
	E_{k} &= \frac{1}{2} \log  \frac{ P+2\rho_{k} \sqrt{P Q_k}+Q_k+N_k } { P'+N_k},\quad k=1,2.\label{eq:Ek_Scalar_Gaussian_Private_Capacity}
	\end{align}
	for some $ \gamma \in [0,1] $ and  $ \rho_1 $, $ \rho_2 $ satisfying $ \rho_1^2+\rho_2^2 \leq 1 $.
\end{theorem}
\begin{IEEEproof}
We start with the converse part, using Proposition \ref{proposition:outer_bound_positive_rates} with $ W = \emptyset $. Define the correlation coefficients between the state and the input sequences as $ 
\rho_{1} \triangleq \frac{\Exp{\rv{X} \rv{S}_{1}}}{\sqrt{PQ_1}} $ and
$ \rho_{2} \triangleq \frac{\Exp{\rv{X} \rv{S}_{2}}}{\sqrt{PQ_2}} $. Define the state variable $ \bv{S} \triangleq (\rv{S}_1,\rv{S}_2) $. And now proceed to lower bound the equivocation measures,
\begin{equation}
I(\bv{S};\rv{Y}_1) = h(\bv{S})-h(\bv{S}|\rv{Y}_1).
\end{equation}
The conditional differential entropy can be upper bounded  as
\begin{equation}
h(\bv{S}|\rv{Y}_1) \leq \frac{1}{2} \log (2\pi e)^2 \frac{Q_1 Q_2 ( P'+1)}{ P+2\rho_1 \sqrt{P Q_1}+Q_1+N_1},
\end{equation}
and
\begin{equation}
h(\bv{S}) =\frac{1}{2} \log (2\pi e)^2 Q_1 Q_2.
\end{equation}
The upper bound on $ E_2 $ follows by similar considerations.

The rates $ R_1 $ and $ R_2 $ can be upper bounded as
\begin{align}
nR_1 &\leq I(\rv{X};\rv{Y}_1|\rv{V},\bv{S}) = h(\rv{X}+\rv{Z}_1|\rv{V},\bv{S})-h(\rv{Z}_1) \label{eq:R1_Gaussian_upper_bound_proof}\\
nR_2 &\leq I(\rv{V};\rv{Y}_2|\bv{S}) = h( \rv{X}+\rv{Z}_2|\bv{S})-h( \rv{X}+\rv{Z}_2|\rv{V},\bv{S}). \label{eq:R2_Gaussian_upper_bound_proof}
\end{align}
The first entropy term in \eqref{eq:R2_Gaussian_upper_bound_proof} can be upper bounded as
\begin{equation}
h( \rv{X}+\rv{Z}_2|\bv{S}) \leq \frac{1}{2} \log (2 \pi e) (P'+N_2).
\end{equation}
Similarly as in Bergmans's proof \cite{bergmans1974} to the converse of Gaussian BC, we first find lower and upper bounds for the second entropy term
\begin{equation}
h( \rv{X}+\rv{Z}_2|\rv{V},\bv{S}) \leq h( \rv{X}+\rv{Z}_2|\bv{S}) \leq \frac{1}{2} \log (2\pi e) ( P'+N_2),
\end{equation}
and
\begin{equation}
h( \rv{X}+\rv{Z}_2|\rv{V},\bv{S}) \geq h(\rv{X}+\rv{Z}_2|\rv{V},\rv{X},\bv{S}) = \frac{1}{2} \log (2 \pi e N_2) .
\end{equation}
Hence, there must exist a $ \gamma \in [0,1] $ such that
\begin{equation} \label{eq:Scalar_Gaussian_Proof_conditional_entropy_equality_gamma}
h( \rv{X}+\rv{Z}_2|\rv{V},\bv{S}) = \frac{1}{2} \log (2\pi e) (  \gamma P'+N_2).
\end{equation}
Now using the conditional EPI, we obtain
\begin{equation}
\begin{split}
h(\rv{X}+\rv{Z}_2|\rv{V},\bv{S}) &= h(\rv{X}+\rv{Z}_1+\tilde{Z}_2|\rv{V},\bv{S}) \\
&\geq  \frac{1}{2} \log \left( 2^{2h(\rv{X}+\rv{Z}_1|\rv{V},\bv{S})}+2^{h(\tilde{Z}_2)} \right) \\
&=  \frac{1}{2} \log \left( 2^{2h(\rv{X}+\rv{Z}_1|\rv{V},\bv{S})}+2 \pi e(N_2-N_1) \right).
\end{split}
\end{equation}
This implies that
\begin{equation} \label{eq:Scalar_Gaussian_Proof_conditional_entropy_upper_bound}
h(\rv{X}+\rv{Z}_1|\rv{V},\bv{S}) \leq \frac{n}{2}\log 2\pi e( \gamma P'+N_1).
\end{equation}
By combining \eqref{eq:R1_Gaussian_upper_bound_proof}, \eqref{eq:R2_Gaussian_upper_bound_proof},  \eqref{eq:Scalar_Gaussian_Proof_conditional_entropy_equality_gamma} and \eqref{eq:Scalar_Gaussian_Proof_conditional_entropy_upper_bound} we have shown that the outer bound on the capacity region consists of rate-pairs satisfying \eqref{eq:R1_Scalar_Gaussian_Private_Capacity} and \eqref{eq:R2_Scalar_Gaussian_Private_Capacity}.

In order to prove the direct part, we use the achievability scheme that was proposed in \cite{steinbergshamai2005}, which integrates Marton coding and Gelfand-Pinsker coding. This scheme was shown to be optimal for Gaussian sources, in the sense that it cancels the state interference completely. In our model $ \bv{S}=(\rv{S}_1,\rv{S}_2) $. We evaluate the mutual information terms in Proposition \ref{proposition:inner_bound_positive_rates} by using the following choice of the auxiliary random variables:
\begin{align}
W &= \emptyset \qquad X'_1\sim \mathcal{N} (0,\gamma P') \qquad X'_2\sim \mathcal{N} (0,\overline{\gamma} P')\\
X &= X'_1+X'_2+\beta_1 \rv{S}_1+\beta_2 \rv{S}_2 \\ 
U &= X'_1+\alpha_{10} X'_2+\alpha_{11} \rv{S}_1 + \alpha_{12} \rv{S}_2 \\
V &= X'_2+\alpha_{21} \rv{S}_1 + \alpha_{22} \rv{S}_2
\end{align}
with $ \beta_1 = \rho_1 \sqrt{\frac{P}{Q_1}} $, $ \beta_1 = \rho_2 \sqrt{\frac{P}{Q_2}} $, $ \alpha_{10} = \frac{\gamma P'}{\gamma P'+N_1} $, $ \alpha_{11} = \frac{(1+\beta_1) \gamma P'}{\gamma P'+N_1} $, $ \alpha_{12} = \frac{\beta_2 \gamma P'}{\gamma P'+N_1} $,
 $ \alpha_{21} = \frac{\beta_1 \overline{\gamma} P'}{P'+N_2} $ and $ \alpha_{22} = \frac{(1+\beta_2) \overline{\gamma} P'}{P'+N_2} $.
Hence,
\begin{gather}
I(\rv{U};\rv{Y}_1)-I(\rv{U};\rv{V},\rv{S}) =  \frac{1}{2} \log \left(1+\frac{\gamma P'}{N_1} \right), \label{eq:Gaussian_inner_bound_R1}\\
I(\rv{V};\rv{Y}_2)-I(\rv{V};\rv{S}) = \frac{1}{2} \log \left(1+\frac{\overline{\gamma} P'}{\gamma P'+N_2} \label{eq:Gaussian_inner_bound_R2} \right).
\end{gather}
The achievability of the equivocation rates follows by showing that
\begin{equation}
I(\rv{S};\rv{U}|\rv{Y}_1)=I(\rv{S};\rv{V}|\rv{Y}_2)=0 .
\end{equation}
Subsituting \eqref{eq:Gaussian_inner_bound_R1} and \eqref{eq:Gaussian_inner_bound_R2} in the equations for $ (R_1,R_2,E_1,E_2) $ we obtain that \eqref{eq:R1_Scalar_Gaussian_Private_Capacity}, \eqref{eq:R2_Scalar_Gaussian_Private_Capacity} and \eqref{eq:Ek_Scalar_Gaussian_Private_Capacity} are achievable and that meets the outer bound  and thus we characterized the achievable region $ \mathcal{R} $ for this channel.
\end{IEEEproof}
\section{Conclusions}
In this paper we addressed the problem of simultaneous communication and state masking over general DMBC with random parameters and parameters given as side information to the encoder. We developed inner and outer bounds on the achievable region containing rates and masking measures and showed that these bounds are tight for the state-dependent Gaussian BC with private messages. Moreover, the standard results as point-to-point masking \cite{merhav2007} and state-dependent BC \cite{steinbergshamai2005} (no masking demands), emerge as special cases of the bounds here. An extension to the MIMO Gaussian BC with private and common messages is under current study.


\section*{Acknowledgment}
The work of M. Dikshtein and S. Shamai (Shitz) has been supported by the European Union's Horizon 2020 Research And Innovation Programme, grant agreement no. 694630 and the Heron consortium via the Israel minister of economy and science.

\appendices

\section{Proof of Proposition \ref{proposition:inner_bound_positive_rates}} \label{section:inner_bound_proof}
Fix the conditional pmf $ P_{\rv{W}\rv{U}\rv{V} \rv{X}|\rv{S}} $.

\paragraph{Codebook generation.} 
Randomly and independently generate $ 2^{n(R_0+\tilde{R}_0)} $ sequences $ w^n(m_0,l_0) $, $ m_0 \in \mathcal{I}_{R_0} $, $ l_0 \in \mathcal{I}_{\tilde{R}_0} $, each according to $ \prod_{i=1}^{n} P_{\rv{W}}(w_i) $. 

For each $ (m_0,l_0) $, randomly and conditionally independently generate $ 2^{n(R_2+\tilde{R}_2)} $ sequences $ v^n(m_0,l_0,m_2,l_2) $, $ m_2 \in \mathcal{I}_{R_2} $, $ l_2\in \mathcal{I}_{\tilde{R}_2} $, each according to $ \prod_{i=1}^n P_{\rv{V}|\rv{W}} (v_i|w_i(m_0,l_0)) $.

Similarly, for each $ (m_0,l_0) $, randomly and conditionally independently generate $ 2^{n(R_1+\tilde{R}_{1s}+\tilde{R}_{12})} $ sequences $u^n(m_0,l_0,m_1,l_{1s},l_{12}) $, $ m_1 \in \mathcal{I}_{R_1} $, $ l_{1s} \in \mathcal{I}_{\tilde{R}_{1s}} $, $ l_{12} \in \mathcal{I}_{\tilde{R}_{12}} $, each according to $ \prod_{i=1}^n P_{\rv{U}|\rv{W}} (u_i|w_i(m_0,l_0))$.

\paragraph{Encoding.} Let $ (m'_0,m'_1,m'_2) $ be the message triple to be sent with the state sequence $ s^n $ observed. In the first encoding step, the encoder finds $ l_0 $ such that
\begin{equation*}
(s^n,w^n(m'_0,l_0)) \in \mathcal{T}_{\epsilon'}^{(n)} (P_{\rv{S}\rv{W}}).
\end{equation*}
If no such $ l_0 $ can be found it declares an error. Let $ \tilde{l}_0 $ specify the chosen $ l_0 $. In the second step, given $ s^n $ and $ w^n(m'_0,\tilde{l}_0)) $ , the encoder finds $ l_2 $ such that
\[ \big(s^n, w^n(m'_0,\tilde{l}_0),v^n(m'_0,\tilde{l}_0,m'_2,l_2) \big) \in \mathcal{T}_{\epsilon''}^{(n)} (P_{\rv{S}\rv{W}\rv{V}}) \]
If no such $ l_2 $ can be found declare an error. Let $ \tilde{l}_2 $ specify the chosen $ l_2 $.  In the third step, fix $ l_{12} $, then, given $ s^n $ and $ w^n(m'_0,\tilde{l}_0) $, the encoder  chooses $ l_{1s} $ such that
\begin{equation*}
(s^n,w^n(m'_0,\tilde{l}_0),u^n(m'_0,\tilde{l}_0,m'_1,l_{1s},l_{12}) \in \mathcal{T}_{\epsilon''}^{(n)} (P_{\rv{S}\rv{W}\rv{U}}).
\end{equation*}
If no such $ l_{1s} $ can be found declare an error. Let $ \tilde{l}_{1s} $ specify the chosen $ l_{1s} $. In the fourth step,  given $ s^n $, $ w^n(m'_0,\tilde{l}_0) $, $ v^n(m'_0,\tilde{l}_0,m'_2,\tilde{l}_2) $ and $ \tilde{l}_{1s} $ the encoder  finds $ l_{12} $ such that
\begin{equation*}
\big(s^n,w^n(m'_0,\tilde{l}_0),v^n(m'_0,\tilde{l}_0,m'_2,\tilde{l}_2),u^n(m'_0,\tilde{l}_0,m'_1,\tilde{l}_{1s},l_{12}) \big) \in \mathcal{T}_{\epsilon'}^{(n)} (P_{\rv{S} \rv{W}\rv{U} \rv{V}}).
\end{equation*}
If no such $ \tilde{l}_{12} $ can be found declare an error. Let $ \tilde{l}_{12} $ specify the chosen $ l_{12} $. Finally, given $ s^n $, $ w^n(m'_0,\tilde{l}_0) $, $ u^n(m'_0,\tilde{l}_0,m'_1,\tilde{l}_{1s},\tilde{l}_{12}) $ and $ v^n(m'_0,\tilde{l}_0,m'_2,\tilde{l}_2) $, generate $ x^n $ with i.i.d. components according to $ \prod_{i=1}^{n} P_{\rv{X}|\rv{S}\rv{W}\rv{U} \rv{V}} (x_i|s_i,w_i,u_i,v_i) $ and convey for transmission.
\paragraph{Decoding.} Upon receiving $ y_1^n $, the decoder at receiver $ 1 $ declares that the pair $ (\hat{m}_0,\hat{m}_1) $ was sent if it is the unique message pair satisfying
\begin{align*}
(w^n(\hat{m}_0,l_0),u^n(\hat{m}_0,l_0,\hat{m}_1,l_{1s},l_{12}),y_1^n) \in \mathcal{T}_\epsilon^{(n)} (P_{\rv{W}\rv{U}\rv{Y}_1})
\end{align*}
for some $ l_0 $, $ l_{1s} $ and $ l_{12} $.

Similarly, Decoder 2 declares that the pair $ (\hat{m}_0,\hat{m}_2) $ was sent if it is the unique message pair such that
\begin{align*}
(w^n(\hat{m}_0,l_0),v^n(\hat{m}_0,l_0,\hat{m}_2,l_2),y_2^n) \in \mathcal{T}_\epsilon^{(n)} (P_{\rv{W}\rv{V}\rv{Y}_2})
\end{align*}
for some $ l_0 $, and $ l_{2} $.

\paragraph{Analysis of the probability of error.} Assume without loss of generality that the message triple $ (\rv{M}_0,\rv{M}_1,\rv{M}_2) = (1,1,1) $ was sent and let $ (L_0,L_{1s},L_{12},L_2) $ be the chosen indexes for $ w^n $, $ u^n $ and $ v^n $. The encoder makes an error only if one or more of the following errors occur:
\begin{align*}
\mathcal{E}_{01} &= \{ (\rv{S}^n,\rv{W}^n(1,l_0))\notin \mathcal{T}_{\epsilon'}^{(n)}(P_{\rv{S}\rv{W}}) \text{ for all } l_0\in \mathcal{I}_{\tilde{R}_0} \}, \\
\mathcal{E}_{02} &= \{ (\rv{S}^n,\rv{W}^n(1,L_0),\rv{V}^n(1,L_0,1,l_2))\notin \mathcal{T}_{\epsilon'}^{(n)}(P_{\rv{S}\rv{W}\rv{V}}) \text{ for all } l_2\in \mathcal{I}_{\tilde{R}_2} \}, \\
\mathcal{E}_{03} &= \{ (\rv{S}^n,\rv{W}^n(1,L_0),\rv{U}^n(1,L_0,1,l_{1s},l_{12})\notin \mathcal{T}_{\epsilon'}^{(n)}(P_{\rv{S}\rv{W}\rv{U}}) \text{ for all } l_{1s}\in \mathcal{I}_{\tilde{R}_{1s}} \}, \\
\mathcal{E}_{04} &= \{ (\rv{S}^n,\rv{W}^n(1,L_0),\rv{V}^n(1,L_0,1,L_2),\rv{U}^n(1,L_0,1,L_{1s},l_{12})\notin \mathcal{T}_{\epsilon'}^{(n)}(P_{\rv{S}\rv{W}\rv{V}\rv{U}}) \text{ for all } l_{12}\in \mathcal{I}_{\tilde{R}_{12}} \}.
\end{align*}
Thus, by the union of events bound, the probability that the encoder at the helper makes an error, can be upper bounded as
\begin{align*}
\Pr(\mathcal{E}_0) &= \Pr (\mathcal{E}_{01} \cup \mathcal{E}_{02} \cup \mathcal{E}_{03} \cup \mathcal{E}_{04}) \\
&\leq \Pr (\mathcal{E}_{01})+\Pr(\mathcal{E}_{01}^c \cap \mathcal{E}_{02}) + \Pr(\mathcal{E}_{01}^c \cap \mathcal{E}_{02}^c \cap \mathcal{E}_{03}) + \Pr(\mathcal{E}_{01}^c \cap \mathcal{E}_{02}^c \cap \mathcal{E}_{03}^c \cap \mathcal{E}_{04})
\end{align*}
By the covering lemma with $ \rv{U} \leftarrow \emptyset $, $ \rv{X} \leftarrow \rv{S} $,  $ \hat{\rv{X}}\leftarrow \rv{U} $, and $ \mathcal{A}=\mathcal{I}_{\tilde{R}_0} $, $ \Pr(\mathcal{E}_{01}) $ tends to zero as $ n\rightarrow \infty $ if $ \tilde{R}_0 > I(\rv{W};\rv{S})+\delta(\epsilon') $.

Similarly, using the covering lemma with $ \rv{U} \leftarrow \rv{W} $, $ \rv{X} \leftarrow \rv{S} $, $ \hat{\rv{X}} \leftarrow \rv{V} $, and $ \mathcal{A} = \mathcal{I}_{\tilde{R}_2} $, $ \Pr(\mathcal{E}_{01}^c \cap \mathcal{E}_{02}) $ tends to zero as $ n\rightarrow \infty $ if $ \tilde{R}_2 > I(\rv{S};\rv{V}|\rv{W}) + \delta(\epsilon') $.

Analogously, using the covering lemma with $ \rv{U} \leftarrow \rv{W} $, $ \rv{X} \leftarrow \rv{S} $, $ \hat{\rv{X}} \leftarrow \rv{U} $, and $ \mathcal{A} = \mathcal{I}_{\tilde{R}_{1s}} $, $ \Pr(\mathcal{E}_{01}^c \cap \mathcal{E}_{02}^c \cap \mathcal{E}_{03}) $ tends to zero as $ n\rightarrow \infty $ if $ \tilde{R}_{1s} > I(\rv{U};\rv{S}|\rv{W}) + \delta(\epsilon') $.

Now consider the probability of the event $ (\mathcal{E}_{01}^c \cap \mathcal{E}_{02}^c \cap \mathcal{E}_{03}^c \cap \mathcal{E}_{04}) $. The triple $ (\rv{W}^n(1,L_0),\rv{U}^n(1,L_0,1,L_{1s},l_{12},\rv{S}^n) $ is jointly typical for every $ l_{12} $, thus
\begin{equation*}
2^{-n(H(\rv{U}|\rv{W},\rv{S})+\delta(\epsilon'))} \leq P_{\rv{U}^n|\rv{W}^n,\rv{S}^n} (u^n(1,L_0,1,L_{1s},l_{12})|w^n(1,L_0),s^n) \leq 2^{-n(H(\rv{U}|\rv{W},\rv{S})-\delta(\epsilon'))}.
\end{equation*}
For any $ l_{12} $, we have
\begin{align*}
&\Pr \{(s^n,w^n,v^n,U^n(1,L_0,1,L_{1s},l_{12})) \in T_{\epsilon''}^{n} | \rv{S}^n=s^n,\rv{W}^n=w^n,\rv{V}^n =v^n \} \\
&=\Pr \{(s^n,w^n,v^n,U^n(1,L_0,1,L_{1s},l_{12})) \in T_{\epsilon''}^{n} | \rv{S}^n=s^n,\rv{W}^n=w^n \} \\
&= \sum_{u^n \in \mathcal{T}^{(n)}_{\epsilon'}(\rv{U}|s^n,w^n,v^n)} P_{\rv{U}^n|\rv{S}^n \rv{W}^n \rv{V}^n} (u^n|s^n,w^n,v^n) \\
&\geq 2^{n(H(\rv{U}|\rv{S},\rv{W},\rv{V})-\delta(\epsilon')} 2^{-n(H(\rv{U}|\rv{S},\rv{W})+\delta(\epsilon')} \\
&\geq 2^{-n(I(\rv{U};\rv{V}|\rv{S},\rv{W})+\delta(\epsilon')}
\end{align*}
\begin{align*}
&\Pr(\mathcal{E}_{01}^c \cap \mathcal{E}_{02}^c \cap \mathcal{E}_{03}^c \cap \mathcal{E}_{04}) = \sum_{(s^n,w^n,v^n) \in \mathcal{T}^{(n)}_{\epsilon'}} P_{\rv{S}^n \rv{W}^n \rv{V}^n}(s^n,w^n,v^n) \\
&\times \Pr \{(s^n,w^n,v^n,U^n(1,L_0,1,L_{1s},l_{12})) \notin T_{\epsilon''}^{n} \text{ for all } l_{12}| \rv{S}^n=s^n,\rv{W}^n=w^n,\rv{V}^n =v^n \} \\
&= \sum_{(s^n,w^n,v^n) \in \mathcal{T}^{(n)}_{\epsilon'}} P_{\rv{S}^n \rv{W}^n \rv{V}^n}(s^n,w^n,v^n) \\
&\times \prod_{l_{12}=1}^{2^{n\tilde{R}_{12}}} \Pr \{(s^n,w^n,v^n,U^n(1,L_0,1,L_{1s},l_{12})) \notin T_{\epsilon''}^{n} | \rv{S}^n=s^n,\rv{W}^n=w^n \} \\
&\leq (1-2^{-n(I(\rv{U};\rv{V}|\rv{S},\rv{W})+\delta(\epsilon'))})^{2^{n\tilde{R}_{12}}} \\
&\leq \exp(-2^{n(\tilde{R}_{12}-I(\rv{U};\rv{V}|\rv{S},\rv{W})-\delta(\epsilon'))})
\end{align*}
which tends to zero as $ n \rightarrow \infty $, provided $ \tilde{R}_{12} \geq I(\rv{U};\rv{V}|\rv{S},\rv{W})+\delta(\epsilon') $.

The decoder at receiver 1 makes an error only if one or more of the following events occur
\begin{align*}
\mathcal{E}_{11} &= \{ (\rv{W}^n(1,L_0),\rv{U}^n(1,L_0,1,L_{1s},L_{12}),\rv{Y}_1^n)\notin \mathcal{T}_{\epsilon}^{(n)}(P_{\rv{W}\rv{U}\rv{Y}_1})  \}, \\
\mathcal{E}_{12} &= \{ (\rv{W}^n(1,L_0),\rv{U}^n(1,L_0,m_1,l_{1s},l_{12}),\rv{Y}_1^n)\in \mathcal{T}_{\epsilon}^{(n)}(P_{\rv{W}\rv{U}\rv{Y}_1}) \text{ for some } m_1 \neq 1 \},\\
\mathcal{E}_{13} &= \{ (\rv{W}^n(m_0,l_0),\rv{U}^n(m_0,l_0,m_1,l_{1s},l_{12}),\rv{Y}_1^n)\in \mathcal{T}_{\epsilon}^{(n)}(P_{\rv{U}\rv{X}_1\rv{Y}_1}) \text{ for some } m_0 \neq 1 \text{ and } m_1 \neq 1 \}.
\end{align*}
Again, by the union of events bound, the probability that the decoder at receiver 1 makes an error, can be upper bounded as
\begin{align*}
\Pr(\mathcal{E}_1) &= \Pr ( \mathcal{E}_{11} \cup \mathcal{E}_{12}\cup \mathcal{E}_{13}) \\
&\leq  \Pr (\mathcal{E}_{0} \cup \mathcal{E}_{11} \cup \mathcal{E}_{12}\cup \mathcal{E}_{13}) \\
&\leq \Pr (\mathcal{E}_{0})+\Pr(\mathcal{E}_{0}^c \cap \mathcal{E}_{11})+\Pr(\mathcal{E}_{0}^c \cap \mathcal{E}_{12})+\Pr(\mathcal{E}_{13})
\end{align*}
We have already shown that $ \Pr (\mathcal{E}_{0}) $ tends to zero as $ n\rightarrow \infty $ if $ \tilde{R}_0 > I(\rv{W};\rv{S})+\delta(\epsilon') $, $ \tilde{R}_2 > I(\rv{V};\rv{S}|\rv{W})+\delta(\epsilon') $, $ \tilde{R}_{1s} > I(\rv{U};\rv{S}|\rv{W})+\delta(\epsilon') $ and $ \tilde{R}_{12} > I(\rv{U};\rv{V}|\rv{S},\rv{W})+\delta(\epsilon') $. Next, note that
\begin{align*}
P_{\rv{Y}_1^n|\rv{S}^n \rv{W}^n \rv{U}^n \rv{V}^n \rv{X}^n(1,1,1)} (y_1^n|s^n,w^n,u^n,v^n,x^n) &= \prod_{i=1}^n P_{\rv{Y}_1|\rv{S}\rv{W}\rv{U}\rv{V}\rv{X}} (y_{1i}|s_i,w_i,u_{i},v_{i},x_i) \\
&= \prod_{i=1}^n P_{\rv{Y}_1|\rv{S}\rv{X}} (y_{1i}|s_i,x_{i})
\end{align*}
Hence, by the conditionally typicality lemma, $ \Pr(\mathcal{E}_{0}^c \cap \mathcal{E}_{11}) $ tends to zero as $ n\rightarrow \infty $.

As for the probability of the event $ (\mathcal{E}_{0}^c \cap \mathcal{E}_{12}) $, $ \rv{U}^n(1,L_0,m_1,l_{1s},l_{12}) $ is pairwise conditionally independent of $ \rv{Y}_1^n $ given $ \rv{W}^n(1,L_0) $, furthermore, $ U^n(1,L_0,m_1,l_{1s},l_{12}) $ is distributed according to $ \prod_{i=1}^{n} P_{\rv{U}|\rv{W}} (u_i|w_i) $. Hence, by the packing lemma, with $ \rv{U} \leftarrow \rv{W} $, $ \rv{X} \leftarrow \rv{U} $, $ \rv{Y} \leftarrow \rv{Y}_1 $ and $ \mathcal{A} =  \mathcal{I}_{R_1} \times \mathcal{I}_{\tilde{R}_{1s}}/L_{1s} \times \mathcal{I}_{\tilde{R}_{12}}/L_{12}   $, $ \Pr (\mathcal{E}_{01}^c \cap \mathcal{E}_{12})  $ tends to zero as $ n\rightarrow \infty $ if $ R_1+\tilde{R}_{1s}+\tilde{R}_{12} < I(\rv{U};\rv{Y}_1|\rv{W}) -\delta(\epsilon) $.

Finally, since for $ m_0 \neq 1 $, $ m_1 \neq 1 $, $ (\rv{W}^n(m_0,l_0),\rv{U}^n(m_0,l_0,m_1,l_{1s},l_{12})) $ is independent of $ (\rv{W}^n(1,L_0),\rv{U}^n(1,L_0,1,L_{1s},L_{12}),\rv{Y}_1^n) $, again by the packing lemma with $ \rv{U} \leftarrow \emptyset $, $ \rv{X} \leftarrow (\rv{W},\rv{U}) $, $ \rv{Y} \leftarrow \rv{Y}_1 $ and $ \mathcal{A}=  \mathcal{I}_{R_0} \times \mathcal{I}_{\tilde{R}_0}/L_0 \times \mathcal{I}_{R_1} \times \mathcal{I}_{\tilde{R}_{1s}}/L_{1s}\times \mathcal{I}_{\tilde{R}_{12}}/L_{12}$, $ \Pr (\mathcal{E}_{13}) $ tends to zero as $ n\rightarrow \infty $ if $ R_0+\tilde{R}_0+R_1+\tilde{R}_{1s}+\tilde{R}_{12} < I(\rv{W},\rv{U};\rv{Y}_1)-\delta(\epsilon) $.

Next consider the average probability of error for decoder 2. The decoder at receiver 2 makes an error only if one or more of the following events occur
\begin{align*}
\mathcal{E}_{21} &= \{ (\rv{W}^n(1,L_0),\rv{V}^n(1,L_0,1,L_2),\rv{Y}_2^n)\notin \mathcal{T}_{\epsilon}^{(n)}(P_{\rv{W}\rv{V}\rv{Y}_2})  \}, \\
\mathcal{E}_{22} &= \{ (\rv{W}^n(1,L_0),\rv{V}^n(1,L_0,m_2,l_2),\rv{Y}_2^n)\in \mathcal{T}_{\epsilon}^{(n)}(P_{\rv{W}\rv{V}\rv{Y}_2}) \text{ for some } m_2 \neq 1 \},\\
\mathcal{E}_{23} &= \{ (\rv{W}^n(m_0,l_0),\rv{V}^n(m_0,l_0,m_2,l_2),\rv{Y}_2^n)\in \mathcal{T}_{\epsilon}^{(n)}(P_{\rv{W}\rv{V}\rv{Y}_2}) \text{ for some } m_0\neq 1 \text{ and } m_2 \neq 1 \}.
\end{align*}
Again, by the union of events bound, the probability that the decoder at receiver 2 makes an error, can be upper bounded as
\begin{align*}
\Pr(\mathcal{E}_2) &= \Pr ( \mathcal{E}_{21} \cup \mathcal{E}_{22}\cup \mathcal{E}_{23}) \\
&\leq  \Pr (\mathcal{E}_{0} \cup \mathcal{E}_{21} \cup \mathcal{E}_{22}\cup \mathcal{E}_{23}) \\
&\leq \Pr (\mathcal{E}_{0})+\Pr(\mathcal{E}_{0}^c \cap \mathcal{E}_{21})+\Pr(\mathcal{E}_{0}^c \cap \mathcal{E}_{22})+\Pr(\mathcal{E}_{23})
\end{align*}
In a very similar fashion as was shown for decoder 1, it can be shown that $ \Pr(\mathcal{E}_2) $ tends to zero as $ n\rightarrow \infty $ if
\begin{align*}
\tilde{R}_0 \geq&\ I(\rv{W};\rv{S})+\delta(\epsilon') \\
\tilde{R}_2 \geq&\ I(\rv{V};\rv{S}|\rv{W})+\delta(\epsilon') \\
\tilde{R}_2+R_2 \leq&\ I(\rv{V};\rv{Y}_2|\rv{W}) -\delta(\epsilon)\\
\tilde{R}_0+R_0+\tilde{R}_2+R_2 \leq&\ I(\rv{W},\rv{V};\rv{Y}_2) - \delta(\epsilon)
\end{align*}
Finally, combining the aforementioned bounds  yields that following region
\begin{align*}
R_0 \leq&\ \min \big\{I(\rv{W};\rv{Y}_1),I(\rv{W};\rv{Y}_2) \big\}-I(\rv{W};\rv{S}) \\
R_2 \leq&\ I(\rv{V};\rv{Y}_2|\rv{W})-I(\rv{V};\rv{S}|\rv{W}) \\
R_1 \leq&\ I(\rv{U};\rv{Y}_1|\rv{W})-I(\rv{U};\rv{V}\rv{S}|\rv{W})
\end{align*}
for some probability distribution $P_{\rv{W}\rv{U}\rv{V}\rv{X}|\rv{S}}$ is achievable. 

As for the upper bound on mutual information between $ \rv{S}^n $ and $ \rv{Y}_2^n $,
\begin{equation} \label{eq:E2_direct_1}
\begin{split}
I(\rv{S}^n;\rv{Y}_2^n) &\leq I(\rv{S}^n;\rv{W}^n,\rv{V}^n,\rv{Y}_2^n) \\
&= I(\rv{S}^n;\rv{W}^n,\rv{V}^n)+I(\rv{S}^n;\rv{Y}_2^n|\rv{W}^n,\rv{V}^n) \\
&= I(\rv{S}^n;\rv{W}^n,\rv{V}^n)+H(\rv{Y}_2^n|\rv{W}^n,\rv{V}^n) -H(\rv{Y}_2^n|\rv{W}^n,\rv{V}^n,\rv{S}^n).
\end{split}
\end{equation}
The first mutual information term can be upper bounded as
\begin{equation} \label{eq:E2_direct_2}
\begin{split}
I(\rv{S}^n;\rv{W}^n,\rv{V}^n) &\leq I(\rv{S}^n;\rv{M}_{0},\rv{M}_2,\rv{W}^n,\rv{V}^n) \\
&\eqann[=]{a}	 I(\rv{S}^n;\rv{W}^n,\rv{V}^n|\rv{M}_0,\rv{M}_2) \\
&= H(\rv{W}^n,\rv{V}^n|\rv{M}_0,\rv{M}_2) - H(\rv{W}^n,\rv{V}^n|\rv{M}_0,\rv{M}_2.\rv{S}^n) \\
&\leq H(\rv{W}^n,\rv{V}^n|\rv{M}_0,\rv{M}_2) \\
&= H(\rv{W}^n|\rv{M}_0,\rv{M}_2)+H(\rv{V}^n|\rv{M}_0,\rv{M}_2,\rv{W}^n) \\
&\eqann[\leq]{b} n(\tilde{R}_0+\tilde{R}_2) \\
&\eqann[=]{c} n(I(\rv{W};\rv{S})+I(\rv{V};\rv{S}|\rv{W})+\epsilon),
\end{split}
\end{equation}
where \eqannref{a} follows since $ (\rv{M}_0,\rv{M}_2,\rv{S}^n) $ are mutually independent and \eqannref{b} due to the fact that the size of each bin of $ \mathcal{C}_0 $ is equal to $ 2^{n\tilde{R}_0} $ and the size of each bin of $ \mathcal{C}_2 $ is equal to $ 2^{n\tilde{R}_2} $. The equality in \eqannref{c} follows from the choice of bin sizes that satisfy error free encoding.
As for the term $H(\rv{Y}_2^n|\rv{W}^n,\rv{V}^n)  $
\begin{equation} \label{eq:conditional_entropy_Y2n_given_WnVn_upper_bound}
\begin{split}
H(\rv{Y}_2^n|\rv{W}^n,\rv{V}^n) &= \sum_{i=1}^{n} H(\rv{Y}_{2,i}|\rv{W}^n,\rv{V}^n,\rv{Y}_2^{i-1}) \\
&\leq \sum_{i=1}^{n} H(\rv{Y}_{2,i}|\rv{W}_i,\rv{V}_i) \\
&= n H(\rv{Y}_{2}|\rv{W},\rv{V}).
\end{split}
\end{equation}
where the inequality follows since conditioning reduces entropy and the last step follows since $ \rv{Y}_{2,i} $ is generated from $ (\rv{W}_i,\rv{V}_i) $ according to
\begin{equation}
\begin{split}
p(y_2|w,v) &= \sum_{s,u,x,y_1} p(s,u,x,y_1,y_2|w,v) \\
&= \sum_{s,u,x,y_1} p(y_1,y_2|s,w,u,v,x) p(s,u,x|w,v) \\
&= \sum_{s,u,x,y_1} p(y_1,y_2|x,s) p(x|w,u,v,s)  p(s,u|w,v).
\end{split}
\end{equation}
where the last equality follows because $ (\rv{W},\rv{U},\rv{V}) \rightarrow (\rv{X},\rv{S}) \rightarrow (\rv{Y}_1,\rv{Y}_2)  $ is a Markov chain.
Next, due to the memorylessness of the channel $ P_{\rv{Y}_1.\rv{Y}_2|\rv{X},\rv{S}} $ and since $ P(x^n|w^n,u^n,v^n) = \prod_{i=1}^{n} P(x_i|w_i,u_i,v_i)$ , we have
\begin{align} 
p(y_2^n|w^n,v^n,s^n ) &= \sum_{u^n,x^n,y_1^n} p(u^n,x^n,y_1^n,y_2^n|w^n,v^n,s^n) \nonumber \\
&\quad= \sum_{u^n,x^n,y_1^n} p(y_1^n,y_2^n|w^n,v^n,x^n,u^n,s^n) p(u^n,x^n|w^n,v^n,s^n) \nonumber \\
&\quad\eqann[=]{a} \sum_{u^n,x^n,y_1^n} p(y_1^n,y_2^n|x^n,s^n) p(x^n|w^n,u^n,v^n,s^n)  p(u^n|w^n,v^n,s^n)  \nonumber \\
&\quad= \sum_{u^n,x^n,y_1^n} \prod_{i=1}^n p(y_{1,i},y_{2,i}|x_i,s_i) p(x_i|w_i,u_i,v_i,s_i) p(u_i|w_i) \nonumber\\
&\quad= \sum_{u^n,x^n,y_1^n} \prod_{i=1}^n p(u_i,x_i,y_{1,i},y_{2,i}|w_i,v_i,s_i) \nonumber \\
&\quad= \prod_{i=1}^n p(y_{2,i}|w_i,v_i,s_i) , \label{eq:p_y1_y2_g_x_s_memorylessness}
\end{align}
where the equality in \eqannref{a} follows because $ (\rv{W},\rv{U},\rv{V}) \rightarrow (\rv{X},\rv{S}) \rightarrow (\rv{Y}_1,\rv{Y}_2)  $ is a Markov chain.
Hence
\begin{equation}  \label{eq:E2_direct_3}
\begin{split}
H(\rv{Y}_2^n|\rv{W}^n,\rv{V}^n,\rv{S}^n) &= \sum_{i=1}^n H(\rv{Y}_{2,i}|\rv{W}_i,\rv{V}_i,\rv{S}_i) \\
&= nH(\rv{Y}_{2}|\rv{W},\rv{V},\rv{S}).
\end{split}
\end{equation}
Finally, by combining \eqref{eq:E2_direct_1}, \eqref{eq:E2_direct_2}, \eqref{eq:conditional_entropy_Y2n_given_WnVn_upper_bound} and \eqref{eq:E2_direct_3} we have
\begin{align*}
I(\rv{S}^n;\rv{Y}_2^n) \leq nI(\rv{S};\rv{W},\rv{V},\rv{Y}_2).
\end{align*}
As for $ I(\rv{S}^n;\rv{Y}_1^n) $, we can similarly upper bound it as
\begin{equation}  \label{eq:E1_direct_1}
\begin{split}
I(\rv{S}^n;\rv{Y}_1^n) &\leq I(\rv{S}^n;\rv{W}^n,\rv{U}^n,\rv{Y}_1^n) \\
&= I(\rv{S}^n;\rv{W}^n,\rv{U}^n)+I(\rv{S}^n;\rv{Y}_1^n|\rv{W}^n,\rv{U}^n) \\
&= I(\rv{S}^n;\rv{W}^n,\rv{U}^n)+H(\rv{Y}_1^n|\rv{W}^n,\rv{U}^n) -H(\rv{Y}_1^n|\rv{S}^n,\rv{W}^n,\rv{U}^n),
\end{split}
\end{equation}
where
\begin{equation} \label{eq:E1_direct_2}
\begin{split}
I(\rv{S}^n;\rv{W}^n,\rv{U}^n) &\leq I(\rv{S}^n;\rv{W}^n,\rv{U}^n,\rv{M}_0,\rv{M}_1 ) \\
&\eqann[=]{a} I(\rv{S}^n;\rv{W}^n,\rv{U}^n|\rv{M}_0,\rv{M}_1) \\
&= H(\rv{W}^n,\rv{U}^n|\rv{M}_0,\rv{M}_1) -H(\rv{W}^n,\rv{U}^n|\rv{M}_0,\rv{M}_1,\rv{S}^n) \\
&= H(\rv{W}^n|\rv{M}_0)+H(\rv{U}^n|\rv{M}_0,\rv{M}_1,\rv{W}^n) \\
&\quad-H(\rv{W}^n|\rv{M}_0,\rv{S}^n)-H(\rv{U}^n|\rv{M}_0,\rv{M}_1,\rv{S}^n,\rv{W}^n) \\
&\eqann[\leq]{b} n\tilde{R}_0+n(\tilde{R}_{1s}+\tilde{R}_{12})-n\tilde{R}_{12} \\
&= n\tilde{R}_0+n\tilde{R}_{1s} \\
&= nI(\rv{W};\rv{S})+nI(\rv{U};\rv{S}|\rv{W}) \\
&= nI(\rv{W},\rv{U};\rv{S}),
\end{split}
\end{equation}
where \eqannref{a} follows since $ (\rv{M}_0,\rv{M}_1,\rv{S}^n) $ are mutually independent and \eqannref{b} is due to the fact that given $ \rv{M}_0 $ there are $ 2^{n\tilde{R}_0} $ sequences $ \rv{W}^n $, similarly, given $ (\rv{M}_0,\rv{M}_1,\rv{W}^n) $, there are $ 2^{n(\tilde{R}_{1s}+\tilde{R}_{12})} $ sequences $ \rv{U}^n $ and given $ (\rv{M}_0,\rv{M}_1,\rv{S}^n,\rv{W}^n) $, there are $ 2^{n\tilde{R}_{12}} $ equiprobable sequences $ \rv{U}^n $. 

By exchanging the roles of $ y_1 $ and $ y_2 $, one can show in similar fashion to \eqref{eq:conditional_entropy_Y2n_given_WnVn_upper_bound} and \eqref{eq:E2_direct_3} that
\begin{equation} \label{eq:E1_direct_3}
H(\rv{Y}_1^n|\rv{W}^n,\rv{U}^n) \leq n H(\rv{Y}_{1}|\rv{W},\rv{U})
\end{equation}
and
\begin{equation} \label{eq:E1_direct_4}
H(\rv{Y}_1^n|\rv{W}^n,\rv{U}^n,\rv{S}^n) = nH(\rv{Y}_{1}|\rv{W},\rv{U},\rv{S}).
\end{equation}
Again, by considering \eqref{eq:E1_direct_1}, \eqref{eq:E1_direct_2}, \eqref{eq:E1_direct_3} and \eqref{eq:E1_direct_4} we have
\begin{equation}
I(\rv{S}^n;\rv{Y}_1^n) \leq nI(\rv{S};\rv{W},\rv{U},\rv{Y}_1) 
\end{equation}
This completes the proof of the inner bound on $ \mathcal{R} $.

\section{Proof of Proposition \ref{proposition:outer_bound_positive_rates}} \label{appendix:outer_bound_common}
	We bound the rates similarly to the Nair-El Gamal proof of the outer bound for general DM-BC with common rates \cite{nair2006}.
\subsection{Common Rate Upper Bound}
By Fano's inequality
\begin{equation} \label{eq:fano_inequality}
H(\rv{M}_0,\rv{M}_k|\rv{Y}_k^n) \leq n(R_0+R_k) P_e^{(n)}+1=n\epsilon_n .
\end{equation}
Observe that
\begin{equation} \label{eq:R0_converse_proof}
\begin{split}
nR_0 &= H(\rv{M}_0) \\
&= I(\rv{M}_0;\rv{Y}_k^n)+H(\rv{M}_0|\rv{Y}_k^n)\\
&\leq I(\rv{M}_0;\rv{Y}_k^n)+H(\rv{M}_0,\rv{M}_k|\rv{Y}_k^n)\\
&\eqann[\leq]{a} I(\rv{M}_0;\rv{Y}_k^n)+n \epsilon_n \\
&= H(\rv{M}_0)-H(\rv{M}_0|\rv{Y}_k^n)+n \epsilon_n \\
&\eqann[=]{b} H(\rv{M}_0|\rv{S}^n)-H(\rv{M}_0|\rv{Y}_k^n)+n \epsilon_n \\
&\leq H(\rv{M}_0|\rv{S}^n)-H(\rv{M}_0|\rv{Y}_k^n,\rv{S}^n)+n \epsilon_n \\
&= I(\rv{M}_0;\rv{Y}_k^n|\rv{S}^n)+n \epsilon_n,
\end{split}
\end{equation}
where \eqannref{a} is due to \eqref{eq:fano_inequality}, \eqannref{b} follows from independency of the state sequence and the common message.

We next proceed to find a single-letter upper bound on $ I(\rv{M}_0;\rv{Y}_k^n|\rv{S}^n) $ for $ k=1,2 $. We start with $ k=1 $.
\begin{equation}
\begin{split}
I(\rv{M}_0;\rv{Y}_1^n|\rv{S}^n) 
&= \sum_{i=1}^{n} I(\rv{M}_0;\rv{Y}_{1,i}|\rv{Y}_{1}^{i-1},\rv{S}^{n}) \\
&\leq \sum_{i=1}^{n} I(\rv{M}_0,\rv{Y}_1^{i-1},\rv{S}^{i-1},\rv{Y}_{2,i+1}^{n},\rv{S}_{i+1}^{n};\rv{Y}_{1,i}|\rv{S_i})  \\
&\leq \sum_{i=1}^{n} I(\rv{W}_{i};\rv{Y}_{1,i}|\rv{S_i})  
\end{split}
\end{equation}
where $ \rv{W}_{i} \triangleq (\rv{M}_0,\rv{Y}_{1}^{i-1}, \rv{S}^{i-1},\rv{Y}_{2,i+1}^n,\rv{S}_{i+1}^n) $. 

In similar fashion we evaluate for $ k=2 $
\begin{equation}
\begin{split}
I(\rv{M}_0;\rv{Y}_2^n|\rv{S}^n) 
&= \sum_{i=1}^{n} I(\rv{M}_0;\rv{Y}_{2,i}|\rv{Y}_{2,i+1}^{n},\rv{S}^{n}) \\
&\leq \sum_{i=1}^{n} I(\rv{M}_0,\rv{Y}_1^{i-1},\rv{S}^{i-1},\rv{Y}_{2,i+1}^{n},\rv{S}_{i+1}^{n};\rv{Y}_{2,i}|\rv{S_i})  \\
&\leq \sum_{i=1}^{n} I(\rv{W}_{i};\rv{Y}_{2,i}|\rv{S_i})  .
\end{split}
\end{equation}

Now consider for $ k \in \{1,2\} $
\begin{equation} \label{eq:R0pRk_converse_proof}
\begin{split}
n(R_0+R_k) &= H(\rv{M}_0)+H(\rv{M}_k) \\
&= I(\rv{M}_0,\rv{M}_k;\rv{Y}_k^n)+H(\rv{M}_0,\rv{M}_k|\rv{Y}_k^n)\\
&\eqann[\leq]{a} I(\rv{M}_0,\rv{M}_k;\rv{Y}_k^n)+n \epsilon_n \\
&= H(\rv{M}_0,\rv{M}_k)-H(\rv{M}_0,\rv{M}_k|\rv{Y}_k^n)+n \epsilon_n \\
&\eqann[=]{b} H(\rv{M}_0,\rv{M}_k|\rv{S}^n)-H(\rv{M}_0,\rv{M}_k|\rv{Y}_k^n)+n \epsilon_n \\
&\leq H(\rv{M}_0,\rv{M}_k|\rv{S}^n)-H(\rv{M}_0,\rv{M}_k|\rv{Y}_k^n,\rv{S}^n)+n \epsilon_n \\
&= I(\rv{M}_0,\rv{M}_k;\rv{Y}_k^n|\rv{S}^n)+n \epsilon_n,
\end{split}
\end{equation}
where, similarly as for the common rate, \eqannref{a} is due to \eqref{eq:fano_inequality}, \eqannref{b} follows from independency of the state sequence and the common message. The mutual information term can be further bounded from above. For $ k=1 $ we have
\begin{equation} \label{eq:R0pR1_converse_proof}
\begin{split}
I(\rv{M}_0,\rv{M}_1;\rv{Y}_1^n|\rv{S}^n) 
&= \sum_{i=1}^{n} I(\rv{M}_0,\rv{M}_1;\rv{Y}_{1,i}|\rv{Y}_{1}^{i-1},\rv{S}^{n}) \\
&\leq \sum_{i=1}^{n} I(\rv{M}_0,\rv{M}_1,\rv{Y}_1^{i-1},\rv{S}^{i-1},\rv{Y}_{2,i+1}^{n},\rv{S}_{i+1}^{n};\rv{Y}_{1,i}|\rv{S_i})  \\
&\leq \sum_{i=1}^{n} I(\rv{W}_{i},\rv{U}_i;\rv{Y}_{1,i}|\rv{S_i})  ,
\end{split}
\end{equation}
where we define the r.v. $ \rv{U}_{i} \triangleq \rv{M}_1 $ for all $ i\in [1:n] $. Similarly it can be shown for $ k=2 $ that
\begin{equation} \label{eq:R0pR2_converse_proof}
\begin{split}
n(R_0+R_2) &\leq  I(\rv{M}_0,\rv{M}_2;\rv{Y}_2^n|\rv{S}^n)+n \epsilon_n \\
&\leq   \sum_{i=1}^{n} I(\rv{W}_{i},\rv{V}_i;\rv{Y}_{2,i}|\rv{S_i}) ,
\end{split}
\end{equation}
where $ \rv{V}_{i} \triangleq \rv{M}_2 $ for all $ i\in [1:n] $.

The rate $ R_1 $ can also be upper bounded as follows
\begin{equation}
\begin{split}
	nR_1 &= H(\rv{M}_1)\\
	&\eqann[=]{a} H(\rv{M}_1|\rv{M}_0)\\
	&= I(\rv{M}_1;\rv{Y}_1^n|\rv{M}_0)+H(\rv{M}_1|\rv{M}_0,\rv{Y}_1^n)\\
	&\leq I(\rv{M}_1;\rv{Y}_1^n|\rv{M}_0)+H(\rv{M}_0,\rv{M}_1|\rv{Y}_1^n)\\
	&\eqann[\leq]{b} I(\rv{M}_1;\rv{Y}_1^n|\rv{M}_0)+n\epsilon_n\\
	&\eqann[=]{c} H(\rv{M}_1|\rv{M}_0,\rv{M}_2,\rv{S}^n)-H(\rv{M}_1|\rv{M}_0,\rv{Y}_1^n)+n\epsilon_n \\
	&\eqann[\leq]{d} H(\rv{M}_1|\rv{M}_0,\rv{M}_2,\rv{S}^n)-H(\rv{M}_1|\rv{M}_0,\rv{M}_2,\rv{S}^n,\rv{Y}_1^n)+n\epsilon_n \\
	&= I(\rv{M}_1;\rv{Y}_1^n|\rv{M}_0,\rv{M}_2,\rv{S}^n)+n\epsilon_n,
\end{split}
\end{equation}
where \eqannref{a} and \eqannref{c} follows from independency of the messages and the state sequence, \eqannref{b} follows from \eqref{eq:fano_inequality} and \eqannref{d} follows from the fact the conditioning decreases entropy.

And in similar fashion
\begin{equation}
nR_2 \leq  I(\rv{M}_2;\rv{Y}_2^n|\rv{M}_0,\rv{M}_1,\rv{S}^n)+n\epsilon_n .
\end{equation}
We use this result to upper bound $ n(R_0+R_1+R_2) $
\begin{equation} \label{eq:Rsum_converse_proof1}
	n(R_0+R_1+R_2) \leq I(\rv{M}_0,\rv{M}_1;\rv{Y}_1^n|\rv{S}^n) + I(\rv{M}_2;\rv{Y}_2^n|\rv{M}_0,\rv{M}_1,\rv{S}^n)+n\epsilon_n.
\end{equation}
We proceed to evaluate the mutual information terms
\begin{align*} 
	&I(\rv{M}_0,\rv{M}_1;\rv{Y}_1^n|\rv{S}^n) + I(\rv{M}_2;\rv{Y}_2^n|\rv{M}_0,\rv{M}_1,\rv{S}^n) \\
	&= \sum_{i=1}^{n} I(\rv{M}_0,\rv{M}_1;\rv{Y}_{1,i}|\rv{Y}_{1}^{i-1},\rv{S}^{n})  + 
	 \sum_{i=1}^{n} I(\rv{M}_2;\rv{Y}_{2,i}|\rv{M}_0,\rv{M}_1,\rv{Y}_{2,i+1}^{n},\rv{S}^n) \\
	&\leq  \sum_{i=1}^{n} I(\rv{M}_0,\rv{M}_1,\rv{Y}_{1}^{i-1};\rv{Y}_{1,i}|\rv{S}^{n}) 	+  \sum_{i=1}^{n} I(\rv{M}_2,\rv{Y}_{1}^{i-1};\rv{Y}_{2,i}|\rv{M}_0,\rv{M}_1,\rv{S}^n,\rv{Y}_{2,i+1}^n) \\
	&=  \sum_{i=1}^{n} I(\rv{M}_0,\rv{M}_1,\rv{Y}_1^{i-1},\rv{Y}_{2,i+1}^{n};\rv{Y}_{1,i}|\rv{S}^{n})+  \sum_{i=1}^{n}  I(\rv{M}_2, \rv{Y}_{1}^{i-1} ;\rv{Y}_{2,i}|\rv{M}_0,\rv{M}_1, \rv{Y}_{2,i+1}^{n},\rv{S}^n) \\
	&\quad  -\sum_{i=1}^{n} I(\rv{Y}_{2,i+1}^{n};\rv{Y}_{1,i}|\rv{M}_0,\rv{M}_1,  \rv{Y}_{1}^{i-1}, \rv{S}^n ) \\
	&=  \sum_{i=1}^{n} I(\rv{M}_0,\rv{M}_1,\rv{Y}_1^{i-1},\rv{Y}_{2,i+1}^n;\rv{Y}_{1,i}|\rv{S}^n)
	+  \sum_{i=1}^{n}  I(\rv{M}_2 ;\rv{Y}_{2,i}|\rv{M}_0,\rv{M}_1, \rv{Y}_{1}^{i-1}, \rv{Y}_{2,i+1}^{n}\rv{S}^n) \\
	& \quad  +  \sum_{i=1}^{n}  I(\rv{Y}_{1}^{i-1} ;\rv{Y}_{2,i}|\rv{M}_0,\rv{M}_1, \rv{Y}_{2,i+1}^{n},\rv{S}^n) 
	-\sum_{i=1}^{n} I(\rv{Y}_{2,i+1}^{n};\rv{Y}_{1,i}|\rv{S}^n,\rv{M}_0,\rv{M}_1,  \rv{Y}_{1}^{i-1} ) .
\end{align*}
Now we apply Csisz\'ar sum identity \cite{csiszar1978broadcast}
\begin{equation}
	\sum_{i=1}^{n} I(\rv{Y}_{2,i+1}^{n};\rv{Y}_{1,i}|\rv{Y}_{1}^{i-1},\rv{S}^{n},\rv{M}_0,\rv{M}_1)
	 = I(\rv{Y}_{1}^{i-1};\rv{Y}_{2,i}|\rv{Y}_{2,i+1}^{n},\rv{S}^n,\rv{M}_0,\rv{M}_1).
\end{equation}
Hence
\begin{equation} \label{eq:Rsum_converse_proof2}
\begin{split}
	&I(\rv{M}_0,\rv{M}_1;\rv{Y}_1^n) \color{black} + I(\rv{M}_2;\rv{Y}_2^n) \\
	&\leq  \sum_{i=1}^{n} I(\rv{M}_0,\rv{M}_1,\rv{Y}_1^{i-1},\rv{Y}_{2,i+1}^n;\rv{Y}_{1,i}|\rv{S}^n)
	+  \sum_{i=1}^{n}  I(\rv{M}_2 ;\rv{Y}_{2,i}|\rv{M}_0,\rv{M}_1, \rv{Y}_{1}^{i-1}, \rv{Y}_{2,i+1}^{n}\rv{S}^n) \\
	&\leq  \sum_{i=1}^{n} I(\rv{M}_0,\rv{M}_1,\rv{Y}_1^{i-1},\rv{S}^{i-1},\rv{Y}_{2,i+1}^n,\rv{S}_{i+1}^n;\rv{Y}_{1,i}|\rv{S}_i)
	+  \sum_{i=1}^{n}  I(\rv{M}_2 ;\rv{Y}_{2,i}|\rv{M}_0,\rv{M}_1, \rv{Y}_{1}^{i-1}, \rv{Y}_{2,i+1}^{n}\rv{S}^n) \\
	&=  \sum_{i=1}^{n} I(\rv{W}_i,\rv{U}_i;\rv{Y}_{1,i}|\rv{S}_i)  +  \sum_{i=1}^{n}  I(\rv{V}_i ;\rv{Y}_{2,i}|\rv{W}_i,\rv{U}_i,\rv{S}_i) .
\end{split}
\end{equation}
	
In a very similar fashion, we can also obtain
\begin{equation} \label{eq:Rsum_converse_proof3}
\begin{split}
n(R_0+R_1+R_2) &\leq  \sum_{i=1}^{n} I(\rv{U}_i;\rv{Y}_{1,i}|\rv{W}_i,\rv{V}_{i},\rv{S}_i)   +  \sum_{i=1}^{n}  I(\rv{W}_i,\rv{V}_{i} ;\rv{Y}_{2,i}|\rv{S}_i) +n\epsilon_n.
\end{split}
\end{equation}

The equivocation bounds can be derived as follows
\begin{equation}
\begin{split}
	I(\rv{S}^n;\rv{Y}_k^n) &= H(\rv{S}^n)-H(\rv{S}^n|\rv{Y}_1^n) \\
	&= \sum_{i=1}^n  H(\rv{S}_i|\rv{S}^{i-1})-H(\rv{S}_i|\rv{S}^{i-1},\rv{Y}_1^n) \\
	&= \sum_{i=1}^n  H(\rv{S}_i)-H(\rv{S}_i|\rv{S}^{i-1},\rv{Y}_1^n) \\
	&\geq \sum_{i=1}^n  H(\rv{S}_i)-H(\rv{S}_i|\rv{Y}_{k,i}) \\
	&= \sum_{i=1}^n I(\rv{S}_i;\rv{Y}_{k,i}).
\end{split}
\end{equation}

Summarizing \eqref{eq:R0_converse_proof}, \eqref{eq:R0pRk_converse_proof}, \eqref{eq:R0pR1_converse_proof}, \eqref{eq:R0pR2_converse_proof}, \eqref{eq:Rsum_converse_proof1}, \eqref{eq:Rsum_converse_proof2} and \eqref{eq:Rsum_converse_proof3}, we have established that
\begin{align*}
	&nR_0 \leq \sum_{i=1}^n I(\rv{W}_i;\rv{Y}_{1,i}|\rv{S}_i) +n\epsilon_n,\\
	&nR_0 \leq \sum_{i=1}^n I(\rv{W}_i;\rv{Y}_{2,i}|\rv{S}_i) +n\epsilon_n,\\
	&n(R_0+R_1) \leq  \sum_{i=1}^{n} I(\rv{W}_{i},\rv{U}_i;\rv{Y}_{1,i}|\rv{S_i}) +n\epsilon_n ,\\
	&n(R_0+R_2)  \leq  \sum_{i=1}^{n} I(\rv{W}_{i},\rv{V}_i;\rv{Y}_{2,i}|\rv{S_i})+n\epsilon_n  ,\\
	&n(R_0+R_1+R_2)  \leq  \sum_{i=1}^{n} I(\rv{W}_i,\rv{U}_i;\rv{Y}_{1,i}|\rv{S}_i)  + \sum_{i=1}^{n}  I(\rv{V}_{i} ;\rv{Y}_{2,i}|\rv{W}_i,\rv{U}_i,\rv{S}_i)+n\epsilon_n ,\\
	&n(R_0+R_1+R_2)  \leq  \sum_{i=1}^{n} I(\rv{U}_i;\rv{Y}_{1,i}|\rv{W}_i,\rv{V}_{i},\rv{S}_i)+  \sum_{i=1}^{n}  I(\rv{W}_i,\rv{V}_{i} ;\rv{Y}_{2,i}|\rv{S}_i) +n\epsilon_n.
\end{align*}
Define the time sharing r.v. $ Q $ to be independent of $ \rv{M}_0 $, $ \rv{M}_1 $, $ \rv{M}_2 $, $ \rv{S}^n $, $ \rv{Y}_1^n $, $ \rv{Y}_2^n $, and uniformly distributed over $ [1:n] $ and define $ \rv{W} \triangleq (\rv{Q},\rv{W}_\rv{Q}) $, $ \rv{U} \triangleq \rv{U}_\rv{Q} $, $ \rv{V} \triangleq \rv{V}_\rv{Q} $, $ \rv{X} \triangleq \rv{X}_\rv{Q} $, $ \rv{Y}_1 \triangleq \rv{Y}_{1,\rv{Q}} $, $ \rv{Y}_2  \triangleq \rv{Y}_{2,\rv{Q}} $ and $ \rv{S} \triangleq \rv{S}_\rv{Q} $.
Hence
\begin{align*}
nR_0 &\leq \sum_{i=1}^n I(\rv{W}_i;\rv{Y}_{1,i}|\rv{S}_i) +n\epsilon_n \\
&= n \sum_{i=1}^n \frac{1}{n} I(\rv{W}_i;\rv{Y}_{1,i}|\rv{S}_i) +n\epsilon_n \\
&= n \sum_{i=1}^n \Prob{\rv{Q}=i} I(\rv{W}_i;\rv{Y}_{1,i}|\rv{S}_i) +n\epsilon_n\\
&= n \sum_{i=1}^n \Prob{\rv{Q}=i} I(\rv{W}_i;\rv{Y}_{1,i}|\rv{S}_i,\rv{Q}=i) +n\epsilon_n ,\\
&= n \sum_{i=1}^n \Prob{\rv{Q}=i} I(\rv{W}_\rv{Q};\rv{Y}_{1,\rv{Q}}|\rv{S}_\rv{Q},\rv{Q}=i) +n\epsilon_n \\
&= n I(\rv{W}_\rv{Q};\rv{Y}_{1,\rv{Q}}|\rv{S}_\rv{Q},\rv{Q}) +n\epsilon_n \\
&\leq  nI(\rv{W};\rv{Y}_1|\rv{S})+n\epsilon_n.
\end{align*}
Similarly
\begin{align*}
&R_0 \leq I(\rv{W};\rv{Y}_{2}|\rv{S}) +\epsilon_n,\\
&R_0+R_1 \leq  I(\rv{W},\rv{U};\rv{Y}_{1}|\rv{S}) +\epsilon_n ,\\
&R_0+R_2  \leq  I(\rv{W},\rv{V};\rv{Y}_{2}|\rv{S})+\epsilon_n,  \\
&R_0+R_1+R_2  \leq  I(\rv{W},\rv{U};\rv{Y}_{1}|\rv{S})  +  I(\rv{V} ;\rv{Y}_{2}|\rv{W},\rv{U},\rv{S})+\epsilon_n ,\\
&R_0+R_1+R_2 \leq I(\rv{U};\rv{Y}_{1}|\rv{W},\rv{V},\rv{S}) + I(\rv{W},\rv{V} ;\rv{Y}_{2}|\rv{S}) +\epsilon_n,
\end{align*}
and
\begin{align*}
	\frac{1}{n} I(\rv{S}^n;\rv{Y}_k^n) \geq I(\rv{S};\rv{Y}_k).
\end{align*}
Furthermore
\begin{align*}
I(\rv{V} ;\rv{Y}_{2}|\rv{W},\rv{U},\rv{S}) &= H(\rv{Y}_{2}|\rv{W},\rv{U},\rv{S})-H(\rv{Y}_{2}|\rv{W},\rv{U},\rv{V},\rv{S}) \\
&\leq H(\rv{Y}_{2}|\rv{W},\rv{U},\rv{S})-H(\rv{Y}_{2}|\rv{W},\rv{U},\rv{V},\rv{X},\rv{S}) \\
&= H(\rv{Y}_{2}|\rv{W},\rv{U},\rv{S})-H(\rv{Y}_{2}|\rv{W},\rv{U},\rv{X},\rv{S}) \\
& = I(\rv{X} ;\rv{Y}_{2}|\rv{W},\rv{U},\rv{S})
\end{align*}
where the last equality is due to the Markov chain $ (\rv{W},\rv{U},\rv{V})\rightarrow \rv{X}\rightarrow (\rv{Y}_1,\rv{Y}_2) $.
likewise
\begin{align*}
	I(\rv{U};\rv{Y}_{1}|\rv{W},\rv{V},\rv{S}) \leq I(\rv{\textsc{X}};\rv{Y}_{1}|\rv{W},\rv{V},\rv{S})
\end{align*}
Since the probability of error is assumed to tend to zero, $ \epsilon_n $ also tend to zero as $ n \rightarrow \infty $.

\bibliography{bibliography}
\bibliographystyle{unsrt}


\end{document}